\documentclass[prb,twocolumn,floatfix]{revtex4}
\usepackage{graphicx}
\usepackage{amsmath}

\begin{document}

\title{Use of dynamical coupling for improved quantum state transfer}

\author{A. O. Lyakhov and C. Bruder}

\affiliation{Department of Physics and Astronomy, University of Basel,
 Klingelbergstr. 82, 4056 Basel, Switzerland}

\begin{abstract}
We propose a method to improve quantum state transfer in transmission
lines.  The idea is to localize the information on the last qubit of a
transmission line, by dynamically varying the coupling constants
between the first and the last pair of qubits. The fidelity of state
transfer is higher then in a chain with fixed coupling constants. The
effect is stable against small fluctuations in the system parameters.
\end{abstract}

\maketitle

\section{Introduction}

Efficient short-distance quantum state transfer is an important
problem in the field of quantum computing. One of the most promising
solutions is to use chains constructed from qubits that are statically coupled
to each other. The idea to use quantum spin chains was initially put
forward by Bose \cite{bose03} and then developed in a number of
papers. These proposals exploit the unitary time evolution governed by
the system Hamiltonian. The state is initialized/encoded at the sender
part of the chain and then, after a certain time, measured/decoded at
the receiving part of the chain. The major advantage of this method is
its simplicity: it does not require controllable coupling constants
between the qubits or complicated gating schemes. It was shown
\cite{bose03} that for short-length chains the fidelity of state
transfer is high, i.e., close to one. But the fact that it is substantially
reduced with the length of the chain triggered the search of methods
that allow to increase the fidelity or even to obtain perfect state
transfer, in the absence of decoherence and relaxation processes.

The main reason for imperfect transfer is the dispersion of the
initial information over the whole chain. Therefore it was proposed to
use spatially varying coupling constants to ``refocus'' the
information at the receiving part of the chain
\cite{christandl,Paternostro,Stolze}. Another possibility is to encode
the information in Gaussian wave packets (with low dispersion) spread
over several spins \cite{Osborne}. Chains where the first and the last
qubits are only weakly coupled to the rest of the chain provide a very
high fidelity \cite{Grudka}, because the intermediate spins are only
slightly excited, which means that dispersion is small. This
method has the major disadvantage that the time required for the
transfer is long compared to the qubit decoherence/relaxation times in
present experimental setups. The idea of so-called conclusive
transfer, providing perfect state transfer using parallel quantum
channels \cite{bose04,Giovannetti04}, is very promising. It can be
realized using almost any spin chain and it is stable against
fluctuations of the chain parameters \cite{bose05}.

Almost all the proposals mentioned above have one common disadvantage: the time
interval for which the fidelity is high is very small for physical qubits and
realistic qubit coupling parameters. For example, for a chain of flux qubits
\cite{ourPaper} with realistic experimental parameters \cite{orlando99}, the
half-width of the first fidelity maximum is about $0.2$ns. At these time scales
state readout and manipulation is impossible using current experimental
technology. Here we show that by dynamically varying the coupling constants
only between the first and the last pair of qubits we can solve this problem
and also increase the fidelity of state transfer.

In real chains the state to be transmitted is initialized in the first
qubit, and this process must not influence the fidelity and dynamics
of the chain. The most natural idea for a full transferring protocol
is as follows: initialize the state in the first qubit, that is decoupled
from the rest of the chain, then adiabatically couple it, wait a certain time
and then adiabatically decouple the last qubit from the chain. This method
requires two controllable gates like one of the proposals for
achieving perfect state transfer \cite{Haselgrove}. In this paper, the
main purpose of the gates is to localize the state on the last qubit
where it can be manipulated during times that are comparable to the
decoherence/relaxation times.

 In the following we use the terms spin and qubit as equivalent. State $|1\rangle$ in
qubit language (which we will also call ``excitation'') corresponds to spin-up
in spin language, and state $|0\rangle$ corresponds to spin-down.

\section{Time-dependent Hamiltonian}

We consider the XXZ-Hamiltonian with time dependent coupling constants between
the first and the last pair of qubits:

\begin{widetext}
\begin{eqnarray}
H(t)=-J_{xy1}(t)(\sigma_{2}^{+}\sigma_{1}^{-}+
\sigma_{2}^{-}\sigma_{1}^{+})-J_{xy}
\sum_{i=3}^{N-1}(\sigma_{i}^{+}\sigma_{i-1}^{-}+
\sigma_{i}^{-}\sigma_{i-1}^{+})\nonumber\\
-J_{xyN}(t)(\sigma_{N}^{+}\sigma_{N-1}^{-}+
\sigma_{N}^{-}\sigma_{N-1}^{+})-J_{z}
\sum_{i=2}^{N}\sigma_{i}^{z}\sigma_{i-1}^{z}- B\sum_{i=1}^{N}\sigma_{i}^{z}\;.
\label{xxz}
\end{eqnarray}
\end{widetext}

This type of Hamiltonian or some of its special cases is used in most of the
papers mentioned above. The XX-part of the Hamiltonian describes the tunneling
of the excitation from one site to another and is a necessary requirement for
quantum state transfer.

\begin{figure}
\includegraphics[width=0.5\textwidth]{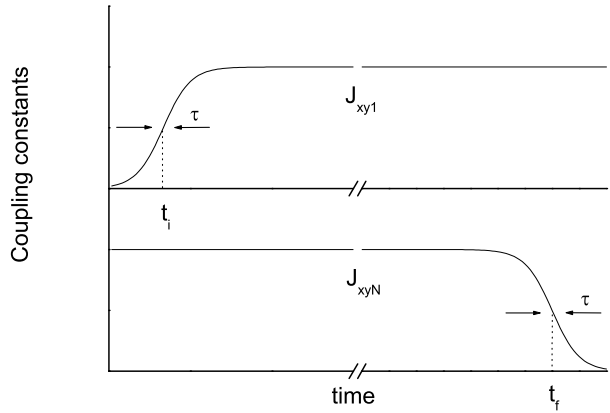}
\caption{Coupling constants $J_{xy1}$ and $J_{xyN}$ as functions of time and
coupling parameters.} \label{fig0a}
\end{figure}

\begin{figure}
\includegraphics[width=0.5\textwidth]{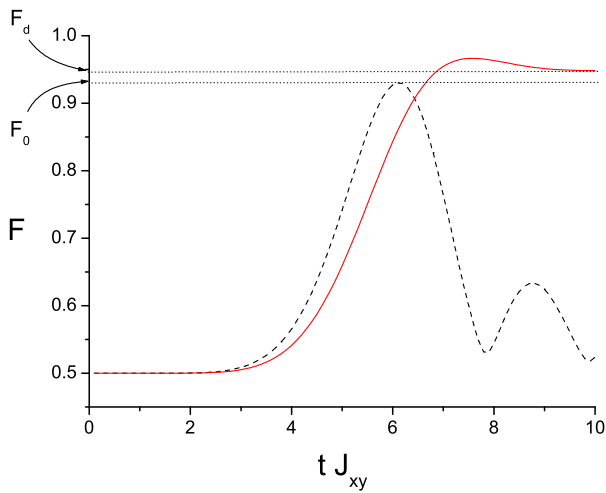}
\caption{Fidelity as a function of time (in units of $J_{xy}^{-1}$) for a chain
with constant coupling parameters (dashed line) and time dependent coupling
parameters (solid line), $N=10$, $t_i=0$, $t_f=6.2/J_{xy}$, $\tau=1/J_{xy}$.}
\label{fig0}
\end{figure}

The physical systems described by this type of Hamiltonian include Josephson
arrays of charge\cite{romito} and persistent-current\cite{ourPaper, levitov}
(flux) qubits, connected by Josephson-junctions/capacitors. The time-dependent
coupling constants can be realized by varying the gate voltages on the
$1$st/$2$nd and $(N-1)$th/$N$th qubits for the flux qubit chain, or by
replacing the Josephson junction between the charge qubits with a SQUID and
varying the flux through it.

As a model we use ``Fermi-function like'' coupling constants:

\begin{equation}
\begin{array}{l}
J_{xy1}(t)=J_{xy}f(t_i,t) \\
J_{xyN}(t)=J_{xy}f(t,t_f)\;,
\end{array}
\label{Eq1}
\end{equation}

with

\begin{equation}
f(t,t^\prime)=\frac{1}{1+\exp{\frac{t-t^\prime}{\tau}}}\;.
\end{equation}

These are smooth functions that vary from $0$ (no coupling) to $J_{xy}$ (full
coupling) and vice versa, see Fig. \ref{fig0a}. The time scale of the
coupling/decoupling procedure is determined by $\tau$. Instant
coupling/decoupling corresponds to $\tau=0$.

Our goal is to calculate the fidelity of the state transfer, the quantity that
characterizes the quality of the transmission line. Let us assume that the
chain is initialized in the state $|00...00\rangle$.  Then, the first qubit is
prepared in the state $|\psi_{in}\rangle$, i.e. the total state of the array is
$|\psi_{in},00...00\rangle$. This is not an eigenstate of the Hamiltonian
(\ref{xxz}), therefore the system will evolve in time. After a time $t$ the
state of the last qubit is read out. Following Bose\cite{bose03}, we average
the fidelity over all pure input states on the Bloch sphere

\begin{equation}
F(t)=\frac{1}{4\pi}\int \langle \psi_{in}|\rho_{out}(t)|\psi_{in}\rangle
\mathrm{d}\Omega
\end{equation}

to obtain a quantity $1/2\leq F(t)\leq 1$ that measures the quality of
transmission independent of $|\psi_{in}\rangle$. Here $\rho_{out}$ is the
reduced density matrix of the last qubit. Fidelity one corresponds to the
perfect state transfer.


By numerically solving the Schr\"odinger equation for the time-dependent
Hamiltonian (\ref{xxz}) we get the fidelity of the state transfer as a function
of time and the coupling parameters $\tau$, $t_i$ and $t_f$. The fidelity has a
complex oscillating behavior. Our goal is to find the coupling parameters that
allow us to localize the state at the last qubit by decoupling it from the rest
of the chain such that the fidelity is maximal. In comparing this fidelity with the static case, we
concentrate on the first maximum: higher maxima appear only after times
much longer than the time at which the first one occurs\cite{ourPaper, romito}. The
typical behavior of $F(t)$ for the static chain in the vicinity
of the first maximum is shown in Fig. \ref{fig0} (dashed line).

\begin{figure}
\includegraphics[width=0.5\textwidth]{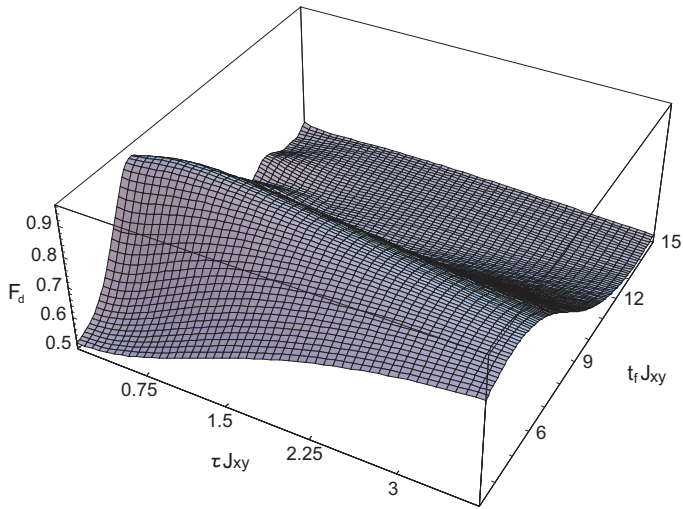}
\caption{Stationary value of the fidelity after decoupling as a function of
$\tau$ and $t_f$, $N=10$, $t_i=0$.} \label{fig1}
\end{figure}

Figure 2 also shows the fidelity in the presence of time-dependent coupling
constants (solid line). One can see that at large time the state is localized at the last
qubit with a fidelity $F_d$ that is higher than for static coupling constants. The time at
which the maximum is achieved is slightly larger. This is natural since in the
presence of the coupling/decoupling procedure the transmission of the
information from the first qubit to the chain and then to the last qubit is
slower. After decoupling, the localized state can be manipulated during a time
interval comparable with the decoherence and relaxation times for the qubit,
which are several orders of magnitude longer then the half-width of the first
fidelity maximum in the static case in present experimental setups. We would
like to mention that the first fidelity maximum in the case of dynamical coupling
constants is even higher than the stationary value of the fidelity after decoupling.
Numerical calculations show that it can exceed the value 0.99 (but, in this case, after
the full decoupling the fidelity will go down to about 0.9).

Figure ~\ref{fig1} shows the fidelity of the state transfer after completely
decoupling the last qubit from the rest of the chain for $t\rightarrow\infty$
as a function of the parameters $\tau$ and $t_f$ (for $t_i=0$). There is a
region where the fidelity for the localized state is higher than in the
time-independent case (up to $4\%$).

\begin{figure}
\includegraphics[width=0.45\textwidth]{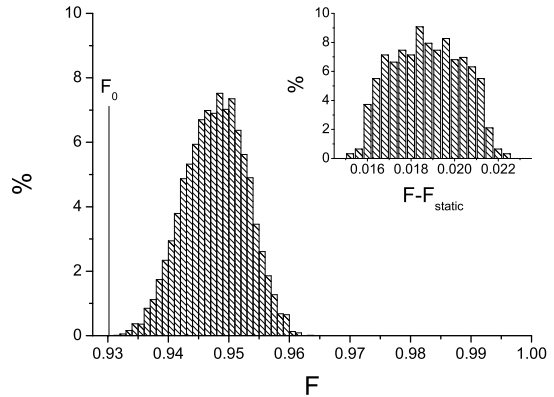}
\caption{Fidelity distribution in the presence of small disorder in the
coupling constants $J_{xy}$, $N=10$, $t_i=0$, $\tau=0.325/J_{xy}$,
$t_f=6.2/J_{xy}$. $F_0$ is the first fidelity maximum for the ideal chain with
static coupling constants. Inset: distribution of the fidelity difference between
the dynamical and statical cases in the presence of equal disorder.}
\label{fig2}
\end{figure}

The origin of this phenomenon is similar to the effect described in Ref.
\onlinecite{Haselgrove}. By dynamically varying the coupling constant between
the first and the second qubit, the information about the state enters the
chain as a wave packet that has small dispersion. This corresponds to some sort
of filtering, an interpretation in agreement with the fact that the
fidelity is higher in the case of equal ``profiles'' for the coupling and
decoupling functions. If we use dynamical decoupling only at the end of the
chain and employ instant coupling to initialize the chain, the maximal possible
fidelity for a chain of $N=10$ qubits drops from about $0.99$ to $0.95$ (but it
is still higher than the fidelity for the time-independent case, which is
around $0.93$).

An intuitive explanation is as follows: during the dynamical decoupling,
the information, that is still dispersed in the chain, will arrive at the last
qubit. Therefore, slow decoupling allows more information to be gathered before
the full decoupling occurs.

Experimental qubit arrays are always inhomogeneous, so in the rest of the paper
we will discuss the effect of static disorder in $J_{xy}$ and dynamical
fluctuations in the coupling/decoupling functions. For charge qubit arrays,
the most important source of inhomogeneity is the variance of the Josephson
energies of the junctions (about $5\%$). In the case of the flux-qubit chain
with capacitive coupling, $J_{xy}$ is a complicated function of the Josephson
and charging energies as well as the capacitance of the coupling capacitor, see
Ref. \onlinecite{orlando99}. A rough estimate using realistic parameters leads
to a variance of $10\%$.

\begin{figure}
\includegraphics[width=0.45\textwidth]{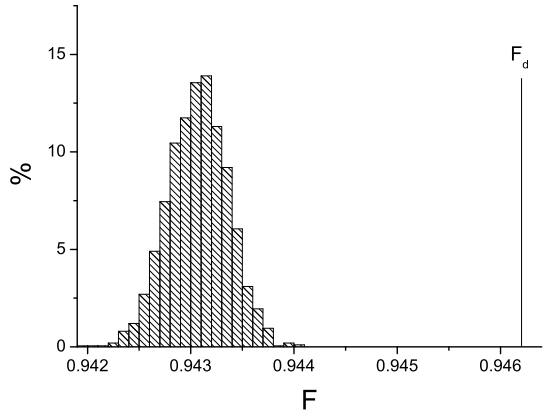}
\caption{Fidelity distribution in the presence of fluctuations in the
coupling/decoupling function, all other coupling constants are fixed and equal.
$t_i=0$, $\tau=0.325/J_{xy}$, $t_f=6.2/J_{xy}$.
$F_{d}$ is the fidelity after decoupling in the absence of fluctuations.} \label{fig4}
\end{figure}

We have performed numerical simulations to evaluate the time evolution of the
system. As a result we find that the phenomena described above, are stable to
static disorder and dynamical fluctuations in the coupling functions, see
Figs.~\ref{fig2},\ref{fig4}. Figure~\ref{fig2} shows the distribution of the
fidelity after complete decoupling in the presence of disorder in the coupling
constants. Its half-width is quite small: even in the worst case the fidelity
is higher than the fidelity of the ideal chain without disorder. The graph was
constructed using a numerical simulation for an ensemble of $10000$ chains
where the coupling constants were of the form $J_{xyi}\rightarrow
J_{xyi}(1+r_i)$, $i=1..N$. The quantity $r_i$ was a random number with uniform
distribution in the interval $[0;0.07]$.

The inset of Fig.~\ref{fig2} shows the difference between the fidelities for
different realizations of the chains with constant and time-dependent
couplings. This difference is around $2\%$, so the effect of increased fidelity
persists. In each realization both chains have the same randomized coupling
constants and the only difference is that $J_{xy1}$ and $J_{xyN}$ are not
multiplied by coupling functions for the time-independent chain.

Figure~\ref{fig4} shows the influence of fluctuations in the
coupling/decoupling functions. Here the coupling constants $J_{xy}$ are
the same for all realizations and the coupling/decoupling functions are of the form

\begin{equation}
\begin{array}{l}
J_{xy1}(t)=J_{xy}\bigg(1+\exp{\frac{t_i-t}{\tau}}\bigg)^{-1}(1+r_1(t))\\
J_{xyN}(t)=J_{xy}\bigg(1+\exp{\frac{t-t_f}{\tau}}\bigg)^{-1}(1+r_N(t))\;.
\end{array}
\label{Eq2}
\end{equation}

The quantities $r_{1,N}(t)$ are stepwise stochastic processes of step
width $0.036\tau$, the step heights are uniformly distributed in the
interval $[0;0.02]$.  One can see that the influence of these
fluctuations is small. We would like to mention, that the fidelity in
the presence of dynamical fluctuations in the coupling functions
is always decreased. This is in agreement
with the filtering idea described above.

\begin{figure}
\includegraphics[width=0.45\textwidth]{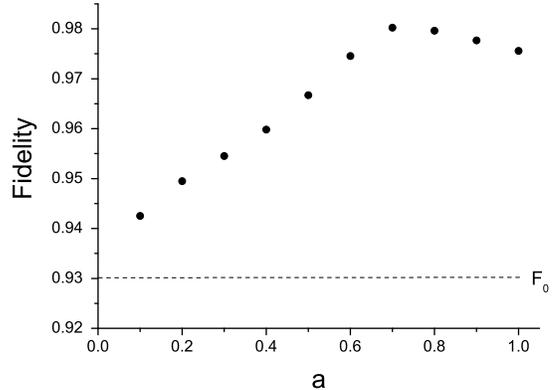}
\caption{Fidelity maxima in the case of coupling functions parameterized as
$J_{xy}((t)/\tau)^a$, $J_{xy}((t_f - t)/\tau + 1)^a$.} \label{fig5}
\end{figure}

Finally, to check that all the effects described above are not the consequence
of our special choice of coupling functions (\ref{Eq1}), we also did the
calculation for another type of dynamical coupling/decoupling:

\begin{equation}
J_{xy1}=\left\{
\begin{array}{ll}
0 & t<0 \\
J_{xy}(t/\tau)^a & t\in[0,\tau] \\
J_{xy} & t>\tau
\end{array} \right.
\end{equation}

\begin{equation}
J_{xyN}=\left\{
\begin{array}{ll}
J_{xy} & t<t_f \\
J_{xy}((t_f - t)/\tau + 1)^a & t\in[t_f,t_f+\tau] \\
0 & t>t_f+\tau
\end{array} \right.
\end{equation}

These functions vary from $0$ to $J_{xy}$ (and vice versa), and we
have chosen $t_i=0$. The parameters $a$ and $\tau$ describe the shape
and timescale of the coupling/decoupling function. The first maxima of
the fidelity for different $a\in[0.1;1]$ are shown in
Fig.~\ref{fig5}. Here, as in Fig.~\ref{fig4}, $\tau$ and $t_f$ are
chosen to maximize it. One can see that this type of dynamical coupling
also allows us to have better state transfer than for the chain with
constant couplings (where the height of the first maximum is $F_0$).
In general, wave packets with bigger width have
lower dispersion.  Therefore we expect that every smooth monotonic
coupling/decoupling function with equal profiles will allow us to
improve the fidelity of state transfer.

\section{Conclusions}

In the past, a number of quantum transmission line systems was
proposed to achieve a perfect or almost perfect state transfer. A
common disadvantage of most of these proposals is the very short time
interval, for which the fidelity of the state transfer is
high. Manipulating the state in such short time intervals is
impossible using current experimental technology. In this paper we
have proposed a method that allows to localize the transferred state
on the last qubit of the transmission line, by varying the coupling
constants between the first and the last pair of qubits. We have also
shown that this method increases the fidelity of the state transfer
and that this effect is stable to static disorder in the coupling
constants and dynamical fluctuations in the coupling/decoupling
functions.

We acknowledge fruitful discussions with S. Bose, D. Burgarth, R. Fazio, and
F.G. Paauw. This work was supported by the European Union under contract
IST-3-015708-IP EuroSQIP, by the Swiss NSF, and the NCCR Nanoscience.

\end{document}